\documentclass[12pt]{iopart}

\usepackage{graphicx}
\usepackage{citesort}

\begin{document}


\title[Shower Power: Isolating the Prompt Atmospheric Neutrino Flux]
{Shower Power: Isolating the Prompt Atmospheric Neutrino Flux Using
Electron Neutrinos}

\author{John F. Beacom}
\address{
Department of Physics, The Ohio State University,
Columbus, OH 43210, USA; \\
Department of Astronomy, The Ohio State University,
Columbus, OH 43210, USA; \\
NASA/Fermilab Astrophysics Center, Fermi National Accelerator Laboratory, 
Batavia, IL 60510-0500, USA \\
Email:  {\tt beacom@mps.ohio-state.edu}}

\author{Juli\'an Candia}
\address{
IFLP/Departamento de F\'{\i}sica, Universidad Nacional de La Plata,
C.C. 67, La Plata 1900, Argentina; \\
Theoretical Physics Department, Fermi National Accelerator Laboratory,
Batavia, IL 60510-0500, USA \\
Email: {\tt candia@fisica.unlp.edu.ar}}

\begin{abstract}
At high energies, the very steep decrease of the conventional
atmospheric component of the neutrino spectrum should allow the
emergence of even small and isotropic components of the total
spectrum, indicative of new physics, provided that they are less
steeply decreasing, as generically expected.  One candidate is the
prompt atmospheric neutrino flux, a probe of cosmic ray composition in
the region of the knee as well as small-$x$ QCD, below the reach of
collider experiments.  A second is the diffuse extragalactic
background due to distant and unresolved AGNs and GRBs, a key test of
the nature of the highest-energy sources in the universe.  Separating
these new physics components from the conventional atmospheric
neutrino flux, as well as from each other, will be very challenging.
We show that the charged-current {\it electron} neutrino ``shower"
channel should be particularly effective for isolating the prompt
atmospheric neutrino flux, and that it is more generally an important
complement to the usually-considered charged-current {\it muon}
neutrino ``track" channel.  These conclusions remain true even for the
low prompt atmospheric neutrino flux predicted in a realistic cosmic
ray scenario with heavy and varying composition across the knee (Candia
and Roulet, 2003 JCAP {\bf 0309}, 005).  We also improve the
corresponding calculation of the neutrino flux induced by cosmic ray
collisions with the interstellar medium.
\end{abstract}

\submitto{Journal of Cosmology and Astroparticle Physics,
{\rm 4 September 2004} \\ 
\phantom{x} \hspace{2cm}  {\rm FERMILAB-PUB-04-188-A-T}}

\maketitle


\section{Introduction}

The measured flux of cosmic rays (CRs) at energies up to about $10^{20}$ eV
reveals the existence of powerful accelerators (or perhaps decaying
supermassive particles), about which very little else is known for
certain \cite{cr04}. Since the directions of cosmic rays can be
scrambled in intervening magnetic fields, point source cosmic ray
astronomy could be difficult to achieve \cite{ha02,dol03,sig04}.  The
same high energy sources may also make gamma rays, which are
directional, but which will be absorbed at high energies and large
distances by the reaction $\gamma\gamma\rightarrow e^+e^-$ on the
cosmic infrared background (e.g., near $10^4$ GeV the mean free path
is about 100 Mpc \cite{kne02,kne04}).  In many models of high energy
sources, neutrinos are also copiously produced.  They have the
advantages of being neither deflected nor absorbed even when traveling
vast distances, and additionally of being able to escape even from
within dense sources.  The obvious disadvantage is that they are
correspondingly hard to detect, due to their only having weak
interactions.

However, for the first time, neutrino telescopes in operation or under
construction will have the required sensitivity to test realistic
models of the highest energy sources in the universe
\cite{le00,al02,hal02,sp03,tor04}.  For example, for several nearby
Active Galactic Nuclei (AGN), high energy gamma rays, up to about
$10^4$ GeV, have been detected~\cite{aha99,kre01,kre02,aha03}.  If
those gamma rays arise from neutral pion decays
($\pi^0\rightarrow\gamma \gamma$), then similar fluxes of neutrinos
from charged pion decays (e.g., $\pi^+\rightarrow\mu^+\nu_\mu$) are
expected.  The pions are naturally produced in models in which a high
energy proton flux collides in the source with either other nucleons
or photons in the ambient radiation field.  The AMANDA detector is
beginning to test these models at a level competitive with gamma ray
telescopes \cite{an01,ah02,ah03a,ah03b,ko03,ac04,ah04}.

Besides point sources, neutrino telescopes can also measure the
diffuse background arising from more distant and higher energy
sources, those which would not be directly visible with gamma rays,
due to the opacity of the cosmic infrared background.  However, it
will be quite challenging to distinguish a diffuse extragalactic
background from the very large flux of neutrinos produced by cosmic
ray collisions with Earth's atmosphere.  
The atmospheric neutrino spectrum falls as $E^{-\gamma}$, 
with the spectral index in the range
$\gamma \simeq 3-3.7$. Due to relativistic time dilation effects, 
the higher the energy of the mesons produced in the atmosphere, the 
larger the amount of energy lost during their propagation before they decay.
Hence, the atmospheric neutrino flux has a spectral index 
similar to the CR spectrum at lower energies (i.e. $\gamma\simeq 3$), 
while it becomes steeper at higher energies. Since the expected  
extragalactic spectrum is harder (indeed, it is thought to fall as $E^{-2}$), 
a non-atmospheric component could be discovered as a break in the spectrum.  
Below the break, the spectrum
would be background dominated, and above the break, signal-dominated.
However, initially the statistics above the break would be poor, both
by definition of a first discovery, and because the spectra are
steeply falling.

The atmospheric neutrinos have been well measured at low energies by
Super-Kamiokande and other detectors~\cite{kaj01,ga02}, and now
have also been detected at higher energies by AMANDA~\cite{wos04}.
The flux seen so far is the ``conventional" atmospheric flux, arising
from pion and kaon decays, and it is reasonably well understood in
terms of the cosmic ray spectrum and composition, meson production
cross sections, and meson propagation and decay in the atmosphere.  
Indeed, the uncertainty for the absolute flux of the low-energy 
atmospheric neutrinos is in the range $10-20\%$ \cite{le00,ga02}.
At higher energies, neutrino fluxes from very short-lived hadrons
dominate, and the ``prompt" atmospheric neutrino flux is much less
understood; empirically, so far not at all.  For these predictions,
there are significant uncertainties due to both the cosmic ray
composition as well as small-$x$ QCD (beyond the reach of colliders);
these issues are discussed in detail below.

So if and when neutrino telescopes first claim discovery of an
extragalactic neutrino flux by a break in the spectrum, the question
will of course arise whether the effect is real, or just a
fluctuation.  In this respect, different detection channels would be
invaluable.  If the signal is real, it could be an important signature
of new physics, in one of two ways: (i) as the prompt atmospheric
neutrino flux, and hence a new probe of both cosmic rays and QCD, or
(ii) as a real extragalactic flux, and hence a new probe of the
high-energy universe.  Distinguishing these possibilities also
requires different detection channels.

The main focus in neutrino telescope studies has been the $\nu_\mu$
charged-current detection channel, to be measured with upward
throughgoing muons. Since by a few hundred GeV, the muon range in ice
exceeds 1 km, the effective detection volume is no longer the detector
volume, but rather the detector area times the muon range, which increases with energy. 
This effect, combined with the rising
neutrino cross section, partially ameliorates the effect of the
steeply falling neutrino spectra.

We propose a new method for isolating the prompt atmospheric neutrino
flux, which, as described above, is important both in its own right,
and as a background to extragalactic fluxes.  Our proposed method
focuses on the channel of $\nu_e$ charged-current events, stressing
the importance of considering the event spectrum as a function of {\it
detectable} energy, and not simply as the product of flux and cross
section as a function of the {\it neutrino} energy.  
Although several analyses based on shower events have already been performed
(for instance, using BAIKAL \cite{ba01,ba02,ay03} and AMANDA \cite{ah03b,ko03,ac04} data), this 
channel has received little attention in the theoretical literature    
\cite{za93,th96,pa98,pa99,vo01,co01,co02,mar03,ca03,ge03,ho03,sta04}. 
However, here we point out that it has several advantages over the usually considered $\nu_\mu$
charged-current detection channel. 
For either the prompt
atmospheric or extragalactic fluxes, the $\nu_e$ fraction is large,
whereas it is small for the conventional atmospheric flux at high
energies.  We will show that in this channel, the spectral break
occurs about an order of magnitude lower in energy than in the
$\nu_\mu$ channel; this is an advantage because at lower energies, the
fluxes are higher and Earth absorption effects are less.  Several
authors have focused on the detection of $\nu_\tau$; however, at
energies below about $5 \times 10^6$ GeV it is challenging to separate
individual $\nu_\tau$ interactions from those of other flavors.  The
$\nu_e$ channel should be particularly effective for prompt
atmospheric neutrinos, since their spectrum falls more steeply than
the extragalactic spectrum, and hence benefits more from a lower
threshold. Moreover, there is much better spectral fidelity between
neutrino energy and detected energy than in $\nu_\mu$ charged-current
interactions, which is important when searching for a spectral break.
It should also be noticed that the intrinsic experimental resolution
of under-ice/water detectors are better for shower events. Indeed,  
the detector response can be better calibrated by means of in-situ 
light sources, and the calorimetric measurement is easier for a shower than for 
a muon track, since in the former the energy is deposited within a small region.  

Below, we present our results in more detail, reviewing the various
fluxes and their characteristics, and how this picture is made more
realistic and in fact more encouraging by considering the {\it
detectable} spectra.  We focus on a realistic prediction for the
prompt atmospheric neutrino flux that takes into account the heavy and
varying cosmic ray composition in the region of the knee~\cite{ca03}.
This model also has important implications for the diffuse neutrinos
from the Galactic center, and we present an improved calculation of
this flux.  We also show how the prompt atmospheric neutrino flux
changes with different assumptions about the cosmic ray composition.
Finally, we summarize our main results.


\section{Calculations and Results}


\subsection{Neutrino Fluxes}

\begin{figure}[t]
\begin{center}
\includegraphics[width=13cm,clip]{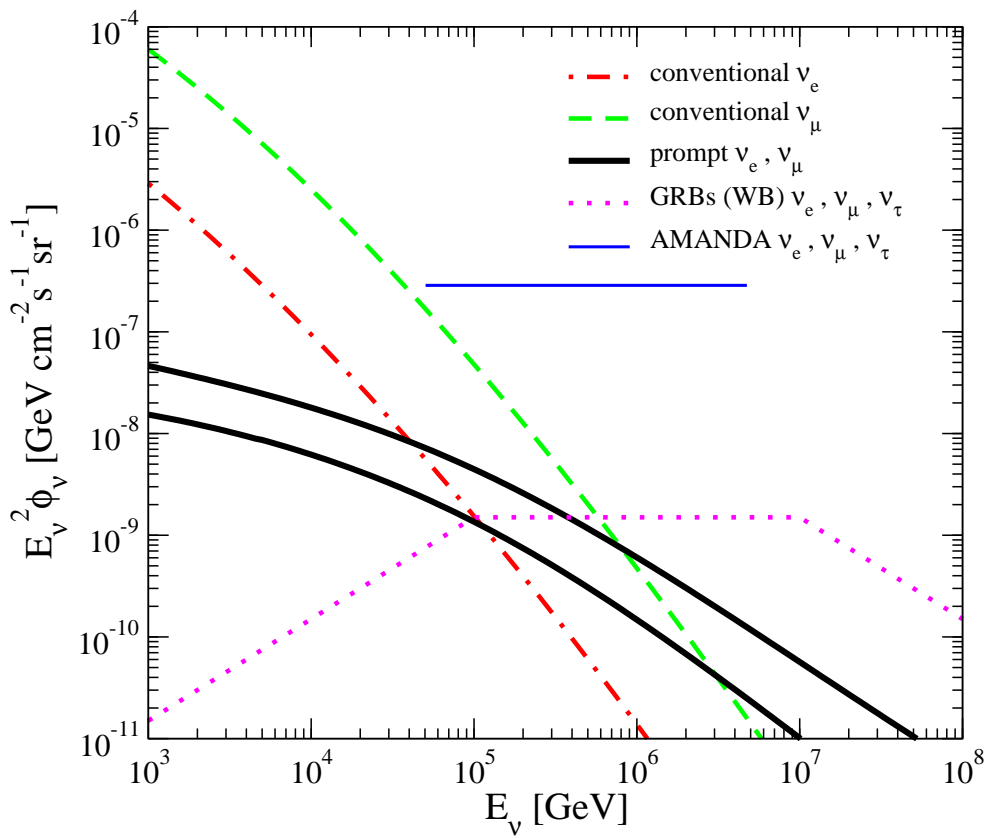}
\caption{\label{fig1} The major components of the high energy neutrino
spectrum are shown, along with labels indicating their flavor content.
Here and throughout, fluxes are given per flavor (but adding neutrinos
and antineutrinos).  For the atmospheric neutrinos, we consider the heavy
and varying cosmic ray composition scenario of Candia and
Roulet~\cite{ca03}; the conventional atmospheric neutrino flux has
been averaged over the zenith angle.  The two lines for the prompt
atmospheric neutrino flux indicate the adopted range of small-$x$ QCD
uncertainties.  As an example of a low diffuse extragalactic flux, the
Waxman-Bahcall prediction for GRBs is shown~\cite{wa99}.  On the high
side, any extragalactic or prompt atmospheric neutrino flux is subject
to the latest AMANDA bound~\cite{ac04}.}
\end{center}
\end{figure}

In Figure~1, we display the main components of the high energy neutrino
flux, and how their relative importance changes across the spectrum.
In Table~I, we list other identifying characteristics of these
components, of which the neutrino flavor ratios are particularly
important.

At low energies, the flux is dominated by conventional atmospheric
neutrinos, which arise from the decays of charged pions and kaons
produced by cosmic rays hitting the top of the atmosphere
\cite{vo80,bu87,bu89,ba89,ho90,li93,ho95,ag96,li98,ba00,fi01,ba03,ho04}.
Although the pion flux is larger than the kaon flux, above a few
hundred GeV, the pions are more likely to interact before decaying,
and due to this the kaon contribution to the neutrino flux dominates
at high energies.  Unlike pions, kaons do decay to electron neutrinos
with an appreciable branching ratio, about $5\%$.  However, this is
small enough to serve as a key part of our argument.  Note that tau
neutrinos arise in the conventional atmospheric neutrino flux only via
neutrino oscillations (very suppressed at high energies), and hence
are ignored here (See Figure~7 of Ref.~\cite{mar03} for an
illustration).

Above about $10^3$ GeV, kaons are also significantly attenuated before
decaying, and the prompt component, arising mainly from the decay of
short lived charmed mesons $D^\pm,D^0,D_s$ and $\Lambda_c$, becomes
increasingly important
\cite{za93,th96,pa98,pa99,vo01,co01,co02,mar03,ca03,ge03,ho03,sta04}.
In these decays, the branching ratios for electron and muon neutrinos
are nearly equal, which is also a key point.  The prompt tau neutrino
flux is about 10 times smaller, and is ignored here.  In Figure~7 of
Ref.~\cite{mar03}, it is shown that the prompt tau neutrino flux
dominates the conventional tau neutrino flux, even above relatively
low energies.  However, it is challenging to individually identify tau
neutrino events in detection until energies of about $5 \times 10^6$
GeV; due to their different propagation in Earth, it may be possible
to recognize their presence in a statistical sense at lower
energies~\cite{ha98,be02,dut02,be03,hus04,yos04,bug04,jo04}.

The evaluation of the prompt neutrino flux requires taking into
account next to leading order processes in the charm production cross
section, which strongly depend on the behavior of the parton
distribution functions at small $x$, below the lowest values ($x \sim
10^{-5}$) probed in collider experiments.  Hence, depending on how the
parton distribution functions are extrapolated, the results appear to
spread over more than an order of magnitude.  In order to illustrate
this uncertainty range, Figure~1 shows results obtained using two
different structure distribution functions, namely the CTEQ3 parton
distribution function \cite{la95,pa99} and the Golec-Biernat,
W\"usthoff (GBW) model \cite{gol99,bar02,mar03,sta04}, which includes
gluon saturation effects.

The prompt atmospheric neutrino flux also strongly depends on the
assumed composition of the cosmic rays. Let
$\phi_Z=\phi_{0Z}E^{-\gamma_Z}$ be the CR spectrum associated with the
CR component of nuclei of charge $Z$ and average mass $A$, where the
spectral index is typically $\gamma_Z\simeq 2.7$ below the knee, and
generally larger above it. This nuclear component provides a nucleon
spectrum $\phi_{N,Z}(E_N)=A^2\phi_Z(E=AE_N)$, which hence corresponds
to a contribution suppressed by a factor $A^{2-\gamma_Z}$ in the
fluxes.  Thus, for the same spectrum, a heavier composition results in
a lower CR nucleon flux, and hence corresponds to lower neutrino
fluxes.  Following Candia and Roulet~\cite{ca03}, a mixed composition
of cosmic rays with all different nuclear species ranging from
hydrogen to nickel was considered.  While the detailed composition of
the different nuclear components below the knee is well known from
experimental observations, at higher energies a rigidity dependent
scenario is assumed, in which each cosmic ray component changes its
spectral index by $\Delta\gamma \simeq 2/3$ across the knee, as can
arise, e.g., in the so-called diffusion/drift
scenario~\cite{can02,can03}.  Below we will discuss the effects of
varying the cosmic ray composition.  In Figure~1, we show both the
conventional and prompt atmospheric neutrino fluxes predicted in this
realistic mixed-composition model of cosmic rays.  While the prompt
atmospheric neutrinos are isotropic, the conventional atmospheric
neutrinos peak at the horizon; in our calculations, we present the
conventional fluxes averaged over the upper hemisphere.

\begin{table}[t]
\begin{center}
\caption{Brief summary of the distinguishing signatures of the
relevant neutrino flux components.  For the different energy spectra,
see the figures.}
\medskip
\begin{tabular}{|l|c|l|}
\hline
Neutrino Flux & Flavors ($ \nu_e : \nu_\mu : \nu_\tau$)
& Angular Dependence \\
\hline
conventional atmospheric & $\frac{1}{20} : 1 : 0$ & peaks at horizon \\
prompt atmospheric & $1 : 1 : \frac{1}{10}$ & isotropic \\
Galactic & $1 : 1 : 1$ & peaks at Galactic center \\
extragalactic & $1 : 1 : 1$ & isotropic; point/transient sources \\
\hline
\end{tabular}
\end{center}
\end{table}

In Figure~1, we also show the latest AMANDA limit on the high-energy
neutrino flux, obtained from their shower analysis~\cite{ko03,ac04}.
The single-flavor AMANDA bound shown in the Figure was obtained neglecting 
single-flavor detection efficiences, and simply dividing by 3 for assumed equal flavor ratios.
Indeed, this bound, which should be regarded as an
estimate for the upper limit of a single-flavor neutrino flux, is also consistent with 
the results of the BAIKAL experiment~\cite{ba01,ba02,ay03}. 
It should be noted that past predictions of the prompt atmospheric
neutrino flux were larger by up to a few orders of magnitude beyond
what we consider here.  While probably not realistic, even a very
large flux would be consistent with the present AMANDA bound.  We
focus on the difficult but realistic case of a small prompt flux.  We
also assume a small extragalactic flux (for illustration, the Waxman
and Bahcall gamma-ray burst (GRB) model~\cite{wa99,ah04}); for a
generic astrophysical neutrino source, one expects a ratio of neutrino
fluxes at production of $1:2:0$, transformed by neutrino oscillations
en route into $1:1:1$ (though new physics in the neutrino sector
could alter both the fluxes and the flavor
ratios~\cite{be03,bea03,bar03,ker03,bea04a,bea04b}).  If the actual
fluxes are larger than assumed here, then our proposed technique will
be easier to use.


\subsection{Detected Spectra}

Figure 1 shows that the prompt atmospheric and the extragalactic
neutrino spectra might not emerge from much larger conventional
atmospheric neutrino spectrum until energies as large as $10^6$ GeV.
To be precise, this is only true for the $\nu_\mu$ spectrum, and the
corresponding charged-current channel based on the detection of
long-ranging muon tracks.  If the $\nu_e$ spectrum could be isolated,
then the spectral break could occur an order of magnitude lower in
energy, where the fluxes are much larger (and note that Figure~1 shows
$E^2 dN/dE$, not the spectra themselves).  Our strategy is therefore
to reduce the conventional atmospheric neutrino background by {\it
excluding} charged-current $\nu_\mu$ events with muon tracks, and
concentrate on $\nu_e$ charged-current events, which initiate showers
(also known as cascades).  As shown in Figure~1 and Table~I, the signals
all have equally large $\nu_e$ and $\nu_\mu$ fluxes, while the
background due to conventional atmospheric neutrinos has a low $\nu_e$
content.  While conventional atmospheric $\nu_\mu$ will contribute to
the shower rate via their neutral-current interactions, their
importance is greatly reduced.

In a neutrino interaction with a nucleon, the neutrino energy $E_\nu$
is shared between the outgoing quark, given a fraction $y$, and the
outgoing lepton, given a fraction $1 - y$.  The differential cross
sections for charged- and neutral-current interactions both peak at $y
= 0$.  In a charged-current $\nu_e$ interaction, the quark initiates a
hadronic shower of energy $\simeq y E_\nu$, and the electron an
electromagnetic shower of energy $\simeq (1 - y) E_\nu$, so that the
total visible energy $E_{vis} \simeq E_\nu$.  We assume that hadronic
and electromagnetic showers are indistinguishable in the detector.  In
a neutral-current interaction, $E_{vis}$ is thus smaller by a factor
$\langle y \rangle \simeq 0.4 - 0.3$ (falling with increasing
energy)~\cite{gan98}.
Further, neutral-current total cross sections are smaller than the
charged-current cross sections, $\sigma_{NC}/\sigma_{CC} \simeq
0.4$~\cite{gan98}.  Taking into account that the conventional
atmospheric neutrino spectrum is very steeply falling (with a
spectral index in the range $\gamma \simeq 3-3.7$, as mentioned above), 
it turns out that the $\nu_\mu$
shower fluxes arising from neutral-current interactions are suppressed
by a factor $\sim \langle y \rangle^{(\gamma-1)} \times
\sigma_{NC}/\sigma_{CC}$, i.e. roughly an order of magnitude relative
to the shower fluxes arising via $\nu_e$ charged-current interactions.

\begin{figure}[t]
\begin{center}
\includegraphics[width=13cm,clip]{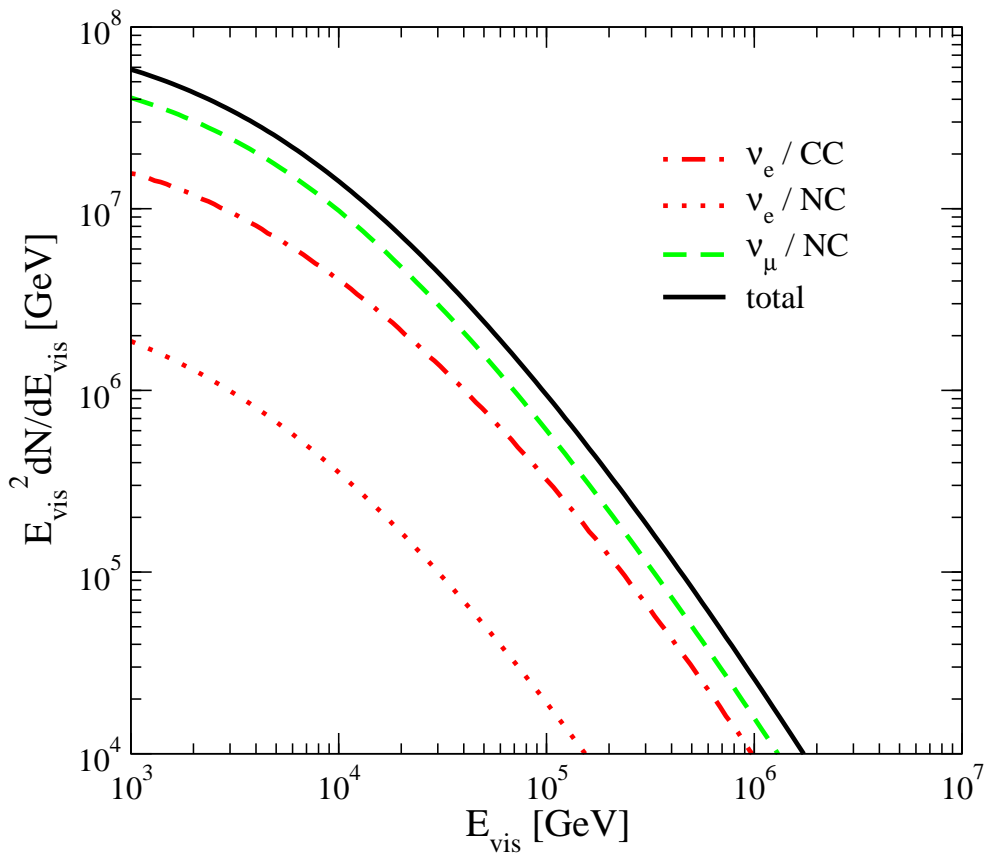}
\caption{\label{fig2} Differential shower rates as a function of
visible energy $E_{vis}$, expected for a km$^3$ detector after 10
years of data taking, using only downgoing neutrinos, and the fluxes
shown in Figure~1.  Only the components of the conventional atmospheric
neutrino spectrum are shown here, with charged-current (CC) and
neutral-current (NC) interactions separated for illustration.  The
relative importance of the $\nu_e$ CC channel grows with respect to
the $\nu_\mu$ NC channel due to the decreasing value of $\langle y
\rangle$.}
\end{center}
\end{figure}

Thus in the detected spectrum of showers from conventional atmospheric
neutrinos, the contributions from $\nu_e$ and $\nu_\mu$ are
comparable, the difference in flux (see Figure~1) being compensated by
the difference in detectability.  Our results for the conventional
atmospheric neutrinos are shown in Figure~2.  As noted, we are excluding
$\nu_\mu$ charged-current events, which can be recognized by the
presence of long-ranging muon tracks.  The spectra shown in the figure
were calculated by convolving the assumed neutrino spectra with the
differential cross section (averaged between neutrinos and
antineutrinos)~\cite{gan98}.  Figure~2 shows that the techniques
described here can greatly reduce the background due to conventional
atmospheric neutrinos.

Since the prompt atmospheric and the extragalactic neutrinos have
equal $\nu_e$ and $\nu_\mu$ fluxes, the corresponding shower rates
will be dominated by $\nu_e$ charged-current interactions.  Though we
include the neutral-current interactions of all relevant flavors, they
could be ignored (e.g., compare the relative $\nu_e$ charged- and
neutral-current rates in Figure~2).  So far, we have not mentioned the
interactions of $\nu_\tau$, should they appear in the flux (see
Table~I).  Below about $E_\nu \simeq 5 \times10^6$~GeV, their
charged-current interactions will produce only showers (with $E_{vis}
\simeq E_\nu$), due to the short lifetime of the tau lepton.  Above
that energy, the length of the tau lepton track becomes long enough
that it could be separated from the shower.  In the conservative
fluxes used in this paper, there is very little flux at such high
energies, and we do not consider the separation of those events.  When
$\nu_\tau$ is present in the flux, we do include their charged- and
neutral-current contributions to the detected shower spectrum.
However, we do note that since the $\nu_\tau$ fraction in the prompt
atmospheric neutrino flux is small, the direct identification of any
$\nu_\tau$ candidates would strongly indicate an extragalactic origin.
A more detailed description of the characteristics of the different
kinds of events in a neutrino telescope, and their relative
detectability, is given in Ref.~\cite{be03}.

\begin{figure}[t]
\begin{center}
\includegraphics[width=13cm,clip]{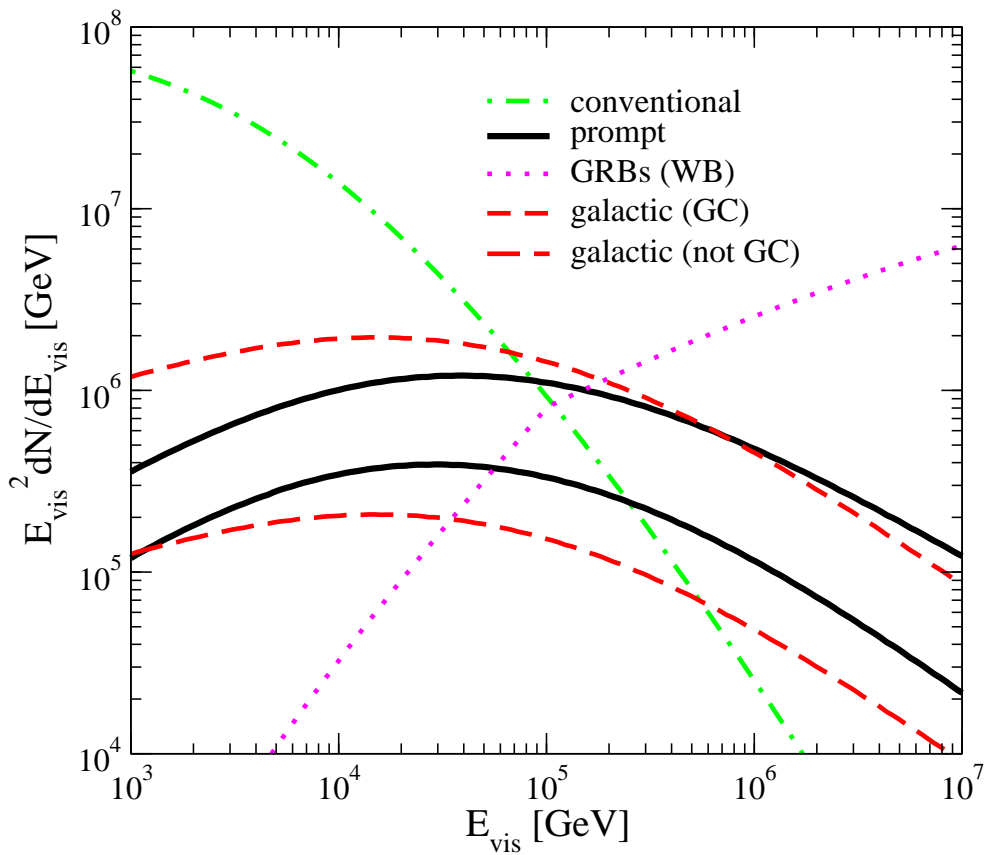}
\caption{\label{fig3} Differential shower spectra as a function of
visible energy $E_{vis}$, expected for a km$^3$ detector after 10
years of data taking, using only downgoing neutrinos, and the spectra
shown in Figure~1.  As explained, we have used a low prompt atmospheric
neutrino flux, corresponding to the realistic cosmic ray composition
model; the AMANDA bound \cite{ac04} would allow it to be about two orders of
magnitude larger.  The spectra from neutrinos produced by cosmic ray
collisions in the galaxy are also shown with long-dashed curves, with
their normalization explained in the text.}
\end{center}
\end{figure}

Figure 3 shows the {\it detectable} shower spectra corresponding to
Figure~1.  The energy at which the prompt atmospheric or extragalactic
signals might emerge from the background of the conventional
atmospheric neutrinos is now an order of magnitude lower.  Had we only
presented the flux or the flux times total cross section versus {\it
neutrino} energy, this important fact would not have been evident.
Hooper et al.~\cite{ho03} also proposed determination of the prompt
atmospheric and the extragalactic neutrino spectra by means of the
shower spectra.  However, there are key differences between our
calculation and theirs.  We assume that all $\nu_\mu$ charged-current
events can be excluded by recognizing their long-ranging muon tracks;
they included $\nu_\mu$ charged-current interactions in the detector
volume, even though they state that it would be better to exclude
them.  More importantly, we expressed the spectra as a function of
{\it visible}, not {\it neutrino} energy, which has a very significant
effect on reducing the background from conventional atmospheric
neutrinos.  Taking these effects into account allows us to
realistically consider much smaller prompt atmospheric and also
extragalactic fluxes than Hooper et al. \cite{ho03}.

The shower channel does not provide specific information on the
neutrino flavor, and it cannot distinguish charged- and
neutral-current events, but this is not a significant disadvantage,
and is more than overcome by the much better fidelity between neutrino
and visible energy, which is essential for resolving a break in the
energy spectrum.  The angular resolution is only moderate ($\simeq
20^\circ$, compared to $\simeq 1^\circ$ for the $\nu_\mu$
charged-current channel), but this is perfectly adequate for an
isotropic flux.

One of the disadvantages of the shower detection channel is that
atmospheric muons can produce a significant background if they pass
near the detector and initiate a shower via a hard bremsstrahlung
event.  Indeed, the current AMANDA shower limit is set completely by
this background~\cite{ko03,ac04}.  However, since this a surface
effect, the much larger size of IceCube should allow reduction of this
background while maintaining a large enough fiducial volume.  
A similar cut on the outer region of the detector will also be necessary
to cut $\nu_\mu$ charged-current events in which the shower registers
in the detector but the muon track escapes. In fact, besides the case 
in which the muon does not emerge from the region of the shower,
the experimental rejection of $\nu_\mu$ charged-current events
also depends on the efficient experimental identification of the muon 
track, which might be relatively dim in comparison to the bright hadronic 
shower. These cuts will reduce the exposure from what we have assumed.
However, it has to be noticed that, in order to avoid the effects
of absorption in Earth, we have considered only downgoing neutrinos.
Neutrinos passing through the whole diameter of Earth are absorbed at
about $4 \times 10^4$ GeV while, for shorter distances, the absorption
energy is significantly higher (e.g., see Figure~2 of
Ref.~\cite{lab04}).
Thus the exposure in the relevant energy range
could be increased from what we assumed by using also a large fraction
of the upgoing events.  Even though it strongly peaks at the horizon,
we have averaged the conventional atmospheric neutrino flux over the
whole upper hemisphere.  A more careful treatment of this background,
cutting events near the horizon, would allow the signals to be seen at
lower energies than shown in our figures.  Finally, as we have
stressed, the prompt atmospheric neutrino flux considered here is low
compared to other models in the literature
\cite{za93,th96,pa98,pa99,vo01,co01,co02,mar03,ca03,ge03,ho03,sta04},
as well as to the current AMANDA limit \cite{ac04}, so there is quite a bit of
room for straightforward improvement of the limit.  A full treatment
of the sensitivity and the ability to separate the flux components,
using the Monte Carlo techniques developed by Kowalski~\cite{ko03},
would be very interesting.


\subsection{Diffuse Galactic Flux}

So far, we have discussed the neutrino fluxes produced either in the
atmosphere or extragalactic sources, but have omitted the diffuse
Galactic flux that arises mainly from pion and muon decays following
cosmic ray interactions with the interstellar
medium\cite{st79,be93,do93,in96,at03} (we neglect the possible
neutrino flux from a point source at the Galactic Center).  In fact,
the diffuse Galactic neutrino flux can be comparable to the other high
energy components \cite{ca03}.  Since the Galactic flux is expected to
be linear in the column depth traversed through the interstellar
medium, it is peaked in the direction of the Galactic center, hence
showing an anisotropy in Galactic coordinates.  In Ref. \cite{be93},
the matter density of the interstellar medium is given as a function
of Galactic coordinates, and a minimum matter density
$n=0.087$~cm$^{-3}$ is assumed.  If the Galactic halo is filled with
this non-negligible matter density, then the anisotropies in the
column density can be ignored except in the direction of the Galactic
Center.  Assuming a halo with a radius of 20 kpc and a vertical scale 
height of 2 kpc, the column density in a
typical direction is $x_{\rm not-GC} \simeq 10^{21}$ cm$^{-2}$;
whereas towards the Galactic Center, $x_{\rm GC} \simeq 10^{22}$
cm$^{-2}$.  
Previously, the Galactic neutrino flux has been estimated in \cite{ca03} 
using $n=1$~cm$^{-3}$ as a representative mean value for the
interstellar matter density in the Galaxy, disregarding the detailed 
dependence of the matter density as a function of Galactic coordinates. 
However, given the large uncertainties in the matter distribution in the Galaxy, 
the estimates of \cite{ca03} for the column density in the directions orthogonal 
and parallel to the Galactic plane are in reasonable 
agreement with the results obtained here. 

We separately consider the contribution from the Galactic
Center region ($|b|\leq 10^\circ$ and $|l|\leq 10^\circ$,
corresponding to a solid angle $\Delta\Omega_{{\rm GC}}=0.12$~sr) and
the averaged contribution from all other directions in the upper
hemisphere (hence corresponding to $\Delta\Omega_{{\rm
not-GC}}=6.16$~sr).  This separation is approximately consistent with
the angular resolution expected for showers in IceCube.

\begin{figure}[t]
\begin{center}
\includegraphics[width=10.5cm,clip]{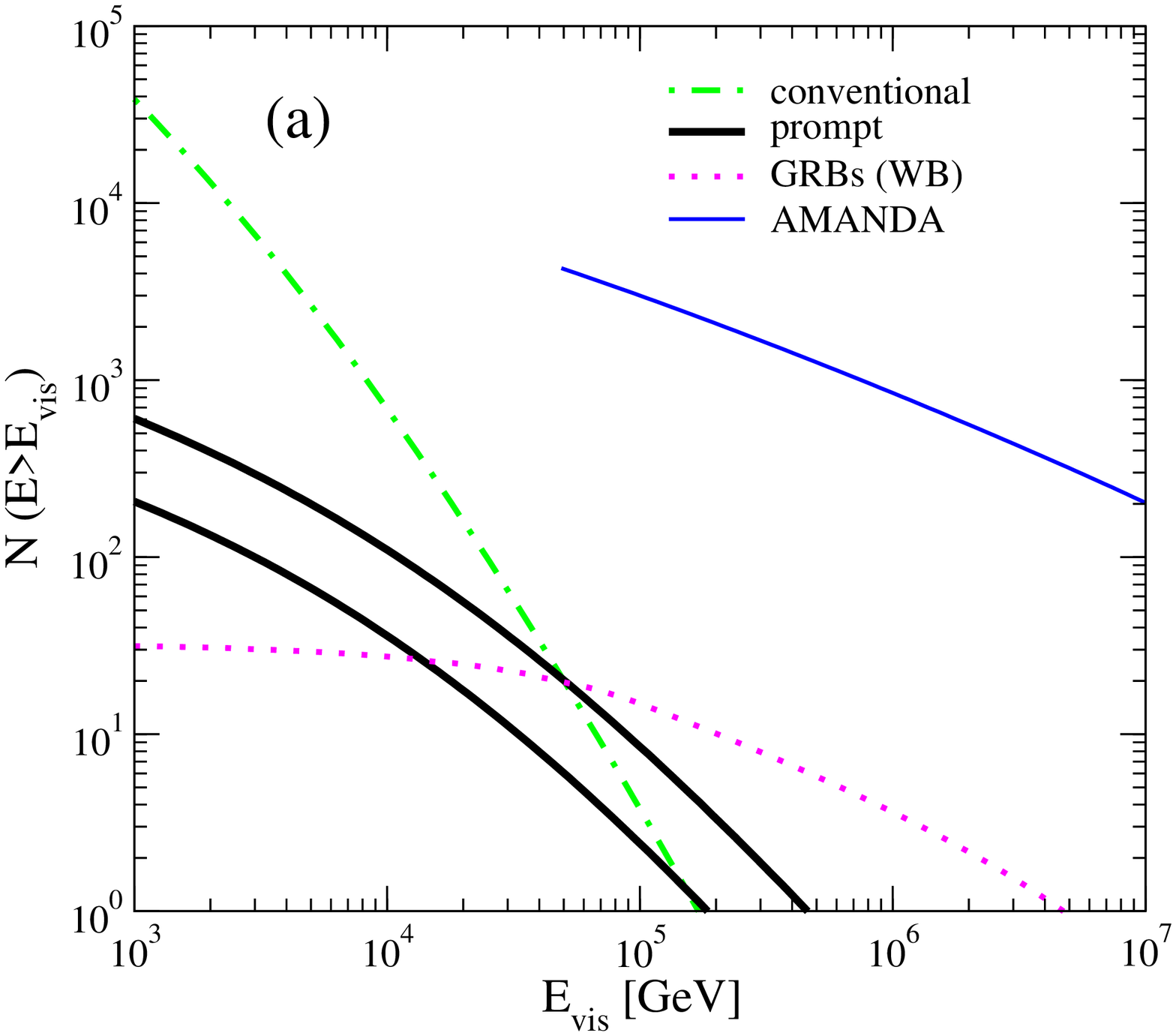}
\includegraphics[width=10.5cm,clip]{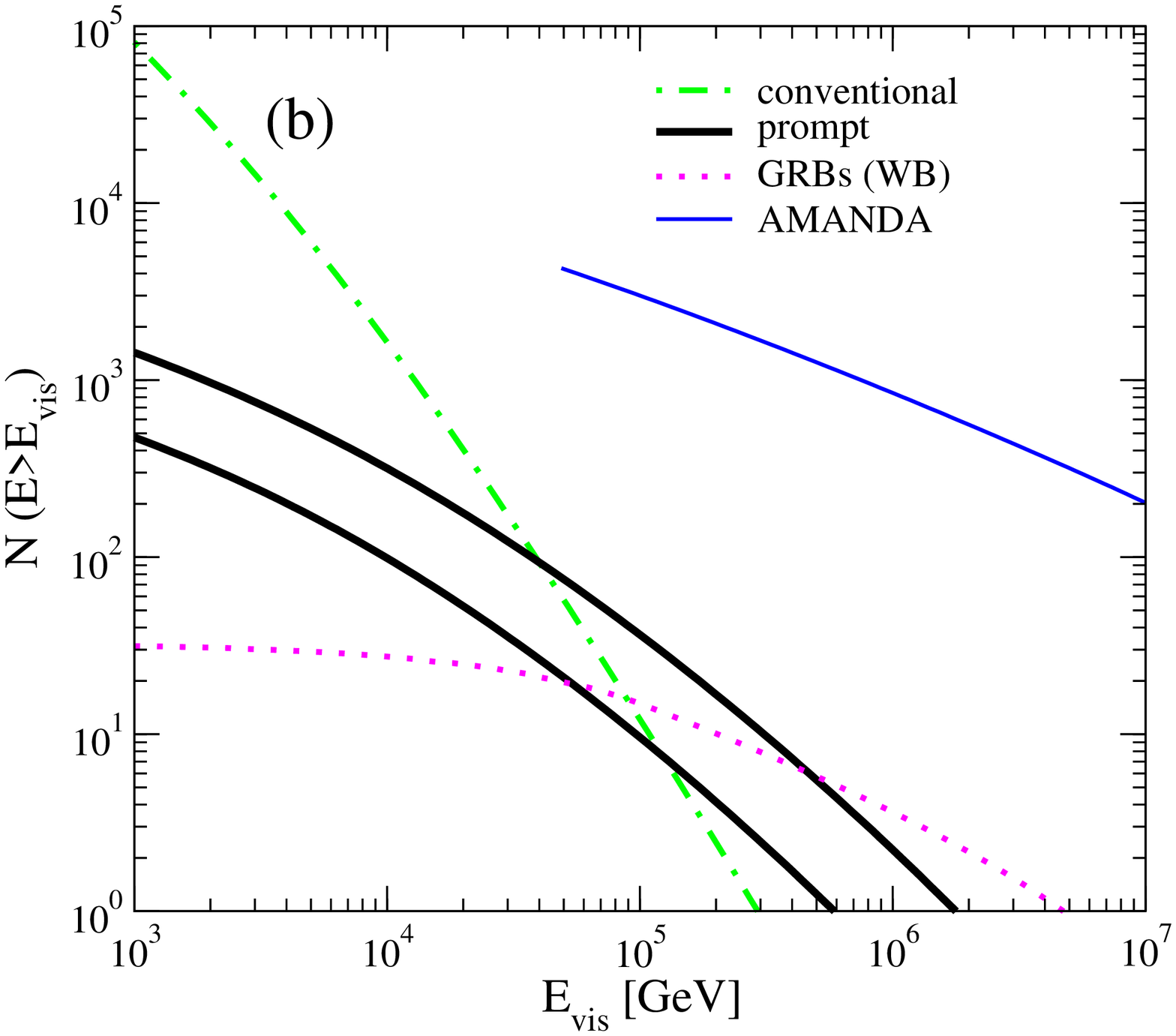}
\caption{\label{fig4} Integral shower rates as a function of visible
energy $E_{vis}$, following Figure 3.  In panel (a), the cosmic ray
spectrum has the heavy and varying composition of Ref.~\cite{ca03}; in
panel (b), the cosmic ray spectrum is assumed to consist only of
protons.  The line marked ``AMANDA" indicates the resulting integral
spectrum assuming an $E^{-2}$ power law, with no upper cutoff,
normalized by the AMANDA differential limit \cite{ac04} (which is actually given
over a slightly smaller energy range).}
\end{center}
\end{figure}

The differential shower rates for the Galactic components are shown in
Figure 3, compared to the fluxes discussed already.  For directions
away from the Galactic Center, the Galactic flux can be neglected
compared to the other components of the spectrum.  In the direction of
the Galactic Center, the Galactic flux is comparable to the other
components.  However, in Figure 3, we have shown the Galactic Center
flux as if it were isotropic.  To calculate the true event rates from
the Galactic Center direction, this and the other fluxes must be
reduced by the small angular acceptance, $\Delta\Omega_{\rm
GC}/\Delta\Omega_{\rm not-GC}\simeq 0.02$, making them too small to be
detectable.


\subsection{Effects of Cosmic Ray Composition}

Figures 4(a)-(b) show the integral shower rates corresponding to the
relevant components of the total high-energy neutrino spectrum.  In
Figure~4(a), it is assumed that all nuclear species in the range
$1\leq Z\leq 28$ contribute to the composition in a rigidity dependent
scenario for the CR knee~\cite{ca03,can02,can03}; in Figure~4(b), it
is assumed that the same CR spectrum is composed only of protons.
Note that the assumed composition affects both the conventional and
prompt atmospheric neutrino fluxes.  As in Figure~1, two lines are
given for the prompt atmospheric neutrino predictions, to indicate the
range of uncertainties arising from different prescriptions for the
small-$x$ QCD.


\section{Conclusions}

By focusing on shower events in which there are no distinguishable
muon tracks, the background coming from the conventional atmospheric
flux is significantly reduced, allowing greater sensitivity to new
physics signals, i.e., the prompt atmospheric neutrino flux or a
diffuse extragalactic neutrino flux coming from unresolved sources.
Considering the shower spectra, the spectral break occurs about an
order of magnitude lower in energy than when considering the usual
$\nu_\mu$ charged-current track channel.  In addition, in the shower
channel there is much closer relationship between neutrino and visible
energy, which will provide better resolution for searching for
spectral break.  This technique should be particularly useful for
measuring the prompt atmospheric neutrino flux; since it is very
steeply falling, it benefits more than an extragalactic neutrino flux
from a reduction in the analysis threshold.  
Although the expected rates, shown
in Figure~4, are not large, they predict the observation of a sufficient 
number of events per year, which make feasible the identification of 
new high energy neutrino signals.  
It should be also noted that the AMANDA bound \cite{ac04} allows larger fluxes, up to about 
two orders of magnitude, which would give much larger rates. Indeed, here we
show that the high energy neutrino signals can be observed even in the 
most pessimistic scenarios assumed for the prompt and extragalactic neutrino fluxes. 
Even before the
prompt atmospheric or extragalactic neutrino fluxes are discovered,
this technique would allow a high statistics measurement of the
conventional atmospheric neutrino flux, essential for verifying its
extrapolation.

Once a break in the spectrum has been observed, several
characteristics can be used to distinguish the prompt atmospheric
neutrino flux from an extragalactic flux.  If there are sufficient
events above the break, the spectra should be quite different.  In
particular, the extragalactic neutrino flux should fall more slowly
and will also initiate more high-energy muons; the one highest-energy
event has a powerful lever arm for determining the spectral
index~\cite{be03}.  The $\nu_\tau$ component is small for the prompt
atmospheric neutrino flux but large for an extragalactic flux; if any
$\nu_\tau$ charged-current events are individually identified at high
energies, then an extragalactic neutrino flux would be
indicated~\cite{be03}.  Finally, the identification of point and/or
transient extragalactic sources will improve the estimates of the
diffuse extragalactic flux due to unresolved sources.

In conclusion, we have shown that the charged-current {\it electron}
neutrino shower channel should be particularly effective for
suppressing the conventional atmospheric neutrino background, leading
to the robust identification of new physics components of the
high-energy neutrino flux, either the prompt atmospheric neutrinos, or
the diffuse extragalactic neutrino background, or both.


\newpage
  
\section{Acknowledgments}

We are grateful to Steve Barwick and Marek Kowalski for discussions. 
We also thank the anonymous referee for their helpful and encouraging remarks. 
J.C. was supported by the Program for Latin American Students of the
Theoretical Physics Department of Fermilab.  This work was supported
by Fermilab (operated by URA under DOE contract DE-AC02-76CH02000), by
NASA grant NAG5-10842, and by funds provided by The Ohio State
University.



\end{document}